\documentclass[]{tMOP2e}
\usepackage{braket,amsfonts,amssymb,graphicx,subfigure,natbib,color,xcolor}

\makeatletter
\renewcommand{\draftnote}{\relax}
\renewcommand{\jname}{\relax}
\renewcommand{\jobtag}{\relax}
\def\articletype#1{\gdef\@articletype{#1}}
\gdef\@articletype{} 
\renewcommand{\ps@plain}{
\renewcommand{\@oddfoot}{\hfil\thepage\hfil}\renewcommand{\@evenfoot}{}
\renewcommand{\@oddhead}{}\renewcommand{\@evenhead}{}}

\makeatother

\begin{document}


\markboth{S.-H.~Tan, L.~A.~Krivitsky and B.-G.~Englert}{Measuring quantum correlations using lossy photon-number-resolving detectors with saturation}


\title{Measuring quantum correlations using lossy photon-number-resolving detectors with saturation}

\author{Si-Hui Tan$^{a,b}$$^{\ast}$\thanks{$^{\ast}$Corresponding author. Email: sihui\_tan@sutd.edu.sg\vspace{6pt}}, Leonid A.~Krivitsky$^{a}$ and Berthold-Georg Englert$^{c,d,e}$\\ \vspace
{6pt}  $^{a}${\em{Data Storage Institute, Agency for Science Technology and Research (A-STAR), Singapore, Singapore}};
$^{b}${\em{Singapore University of Technology and Design, Singapore, Singapore}};
$^{c}${\em{Centre for Quantum Technologies, National University of Singapore, Singapore, Singapore}};
$^{d}${\em{Department of Physics, National University of Singapore, Singapore, Singapore}};
$^{e}${\em{MajuLab, CNRS-UNS-NUS-NTU International Joint Unit, UMI 3654, Singapore, Singapore}}
\\ \vspace{6pt}}

\maketitle

\begin{abstract}
The variance of difference of photocounts (VDPs) is an established measure of quantum correlations for quantum states of light. It enables us to discriminate between the classical correlation of a two-mode coherent state and the quantum correlation of a twin-beam state. We study the effect of loss and saturation of the photon-number-resolving detector on the measurement of the VDPs. An analytic function is derived for this variance, both for the coherent and the twin-beam states. It is found that the VDPs is no longer a reliable entanglement measure in the nonlinear regime of the detector response but it remains useful in some range of values of average photon numbers of the incident light. We also quantify the linear regime of the detector with saturation which will be useful for calibration of the detector quantum efficiency.\bigskip

\bigskip

\end{abstract}

\section{Introduction}

Generation and characterization of multiphoton entangled states represent one of the prerogatives of modern quantum optics experiments. Highly-entangled multiphoton states such as NOON states \cite{noon} and photon-number-correlated twin-beam states \cite{pnc} might become promising resources for practical applications in quantum lithography \cite{litho}, quantum metrology, \cite{metrology} and quantum cryptography \cite{cryptography}. Measurement of the degree of quantum correlations is an essential task for characterizing states of these types. The recent emergence of photon-number-resolving detectors (PNRDs) \cite{migdall, PNRD}, such as visible-light photon counters (VLPCs) \cite{cryo1}, transition-edge sensors (TESs) \cite{cryo2}, time-multiplexed detectors (TMDs) \cite{TMD} and multipixel photon counters (MPPCs) \cite{Silberberg}, makes the direct characterization of multiphoton states by photocounting possible.

A measurement using a $N$-photon-resolving detector can be modeled by a positive-operator valued measure (POVM) with $N+1$ outcomes: $\{\Pi_0, \Pi_1,\ldots, \Pi_N\}$, satisfying completeness, $\sum_{i=0}^N \Pi_i =I$, and positivity, $\Pi_i\geq0$, for all $i=0,\ldots, N$. The outcomes of the POVM are diagonal in the Fock-state basis and hence phase insensitive. The probability of detecting $i$ photons is $p_i'={\rm tr}(\rho\Pi_i)$, where the statistical operator $\rho$ 
describes the photon state, and knowledge of the $p_i'$s enables us to reconstruct the photon statistics. For a perfect detector with no loss (unit quantum efficiency) and no saturation, $\{\Pi_i=\ket{i}\bra{i}, i=1,2,\ldots\}$. Then the probabilities $p_i'$ should reflect the true frequency of occurrence of $i$ 
photons received, $p_i$, in the limit of a large number of detection trials. However, because of loss and noise in the detector the measurement is a proper POVM rather than a von Neumann measurement and the measured probability $p_i'$ is a mixture of the actual photon statistics,
\begin{eqnarray}\label{cond}
p_i'=\sum_j w_{ij} p_j \ ,
\end{eqnarray}
where $w_{ij}$ is the conditional probability of detecting $i$ photons when $j$ photons impinged on the detector. The conditional probabilities model the measurement 
process in the detector; in the POVM picture, they are related to the outcomes by 
$w_{ij}=\bra{j}\Pi_i\ket{j}$, where $\ket{j}$ is the ket for the Fock 
state with $j$ photons. The set of $w_{ij}$ has been used to quantify the capability of photon number discrimination of the PNRD \cite{statistics1}, and the objective of detector tomography is to determine the conditional probabilities $w_{ij}$ in a model-independent way \cite{dt1, 
dt2, dt3}. However, detector tomography relies on the ability to prepare a tomographically complete set of states and complicated optimization techniques. Therefore, parameter estimation for instrument calibration requires suitable models for the 
POVM \cite{Silberberg, Ramilli, Kalashnikov, Gisin, Kok, SVA2012}.

Highly entangled states can be produced in various nonlinear optical processes 
\cite{intense}. At high intensities, the photodetector response becomes nonlinear due 
to saturation. For example, in a silicon multi-pixel photodetector consisting of an  
array of avalanche photodiodes (APD), saturation of individual APDs results in the saturation of the array. TES is another alternative for PNRD, which exploits bolometric methods for photon detection. TESs operate at cryogenic temperatures (below 100mK) and have high quantum efficiency ($\sim 95\%$). Their saturation at high intensities is due to the heating of the detectors beyond the superconducting state. Hence, there is a clear need to understand the effects of saturation so that one can use these PNRDs in experiments with intense non-classical states of light.

In this paper, we assess the feasibility of using lossy PNRDs for measuring quantum correlations under joint photodetection in the presence of saturation. The figure of merit that we use here is the variance of difference of photocounts (VDPs), which has been shown to be a good measure of entanglement \cite{jpdparis}. The dependence of the VDP on quantum efficiencies of photodetectors has been exploited in the method of absolute calibration of quantum efficiency using twin-beam light \cite{Klyshko, cali, Agafonov_cal}. The effects of saturation on this calibration protocol will be discussed here too.

This paper is organized as follows: In Section 2, the detection setup is described. The POVM modeling of losses and saturation is presented. The first and second moments of the photocounts of these POVM for a state with Poissonian statistics are derived. In Section 3, the VDP and its moment operators are described. The VDP of two pairs of two-mode states, one with classical and another with quantum correlations, are calculated. The difference of these VDP, $Q$, is introduced as a measure of discrimination between quantum and classical correlations. A tractable analytic form is derived for $Q$. In Section 4, we introduce the noise reduction factor (NRF), a normalized form of the VDP, and show how an absolute calibration of two detectors can be performed using the NRF in the linear regime of the detector response. The analytic form of the photocount calculated in Section 2 gives a quantitative estimate for the range of mean photon numbers in which the detector remains linear. In Section 5, the results are summarized and the use and limitations of the measurement model are discussed.

\section{Modeling photon detection with loss and saturation}
 
The joint measurement scheme is as follows: Two modes of radiation with the mode creation operators, $a_1$ and $a_2$, are incident on two PNRDs. The resulting photocounts of the respective PNRDs are analyzed. The joint detection of quantum correlations with lossy photodetectors has been looked into in \cite{jpdparis}. Here we extend the theory to include the effects of saturation. For the purpose of the analysis, we use the twin-beam (TWB) state defined as
\begin{eqnarray}\label{twb}
\ket{X}=\sum_{n=0}^\infty b_n \ket{n}_1\ket{n}_2 \ ,
\end{eqnarray}
where $\ket{n}_j, j=1,2$ is the Fock state with $n$ photons in mode $j$ and $b_n$ is the probability amplitude for $\ket{n}_1\ket{n}_2$ in the state $\ket{X}$. It follows from the structure of Eqn. (\ref{twb}) that the two modes of the TWB state are perfectly correlated in photon numbers. The TWB state can be generated by spontaneous parametric down-conversion (SPDC) in a nonlinear crystal pumped by a pulsed laser. The two beams created by SPDC comprise inherently of multiple frequency modes. In the limit of a large number of these modes, $|b_n|^2=\exp(-\bar{n})\bar{n}^n/n!$ applies, where $\bar{n}$ is the mean number of photons in either the signal or the idler mode \cite{waks, haderka}. As a standard for comparison with classical states, we shall use the two-mode coherent (TMC) state $\ket{\alpha}_1\ket{\alpha}_2$ where $\bar{n}=|\alpha|^2$. Without loss of generality, we can assume that there is no phase difference between the two modes in the TMC state because photocounting is a phase-insensitive measurement.

The $m$-th outcome of the POVM of a lossy detector with quantum efficiency $\eta
$ is given by \cite{GaussianCV},
\begin{eqnarray}\label{lossy}
\Pi_m=\sum^\infty_{n=m}w_{m,n}(\eta)|n\rangle \langle n| \ ,
\end{eqnarray}
with
\begin{eqnarray}\label{lossy2}
w_{m,n}(\eta)=\eta^m (1-\eta)^{n-m}\left(\begin{array}{c}n \\ m \end{array}
\right) \ ,
\end{eqnarray}
for $m=0, 1, 2 \ldots$ and $w_{m,n}(\eta)$ was introduced in Eqn. (\ref{cond}). For a PNRD, the measured number of photons, $m$, which is an eigenvalue of the 
photocount, $\widehat{m}$, has an upper bound at $N$. When $m<N$, then its POVM 
is the same as in Eqn. (\ref{lossy}), but because the PNRD is unable to resolve 
between $N$ and $N+1$ or more photons, we have
\begin{eqnarray}\label{N}
\Pi_N=I - \sum_{m=0}^{N-1}\Pi_m \ ,
\end{eqnarray}
for the $N$-th outcome, where $I$ is the identity operator and $N$ is the maximum resolvable photocount. If $\eta<1$ then more than $N$ incident photons will register $N$ or fewer counts. When $N=1$, this POVM reduces to that of the ON/OFF photodetector: $\{\Pi_0,\Pi_1\}$, where $\Pi_0 = \sum_{n=0}^\infty (1-\eta)^n \ket{n}\bra{n}$ and $\Pi_1 = I-\Pi_0$.

The POVM given in Eqn.~(3) has been successfully used to represent PNRDs such 
as the TES \cite{supercondcal} and silicon multi-pixel photodetector 
\cite{KT2012} (where in the latter, the effects of crosstalk have to also be accounted for). Other POVMs exist that are suitable for modeling other types of \\ PNRDs. Here, for comparison, the Sperling-Vogel-Agarwal model \cite{SVA2012} for multiplexed PNRDs is investigated numerically. For $N>1$ and $m=0,1,\ldots, N$, the $m$-th outcome of this model is given by
\begin{align}\label{SVmodel}
\Pi_m^{\rm(sv)}=:\frac{N!}{k!(N-k)!}(e^{-\eta\frac{\hat{n}}{N}})^{N-k}(I-e^{-\eta\frac{\hat{n}}{N}})^k: \ ,
\end{align}
where $:\ :$ denotes normal ordering of the creation and annihilation operators. The two models given, the first given by Eqns.~(\ref{lossy})--(\ref{N}), and the second by Eqn.~(\ref{SVmodel}) are two mathematical descriptions of detectors that operate under different physical conditions. However, in the next section we will show that these POVMs do not lead to vastly different behavior. Thus, we shall proceed with our analysis using only the POVM given by Eqns.~(\ref{lossy})--(\ref{N}) in the rest of the paper.

\subsection{Analytic expressions for the mean of the first- and second-moment operators of $\{\Pi_m\}$ for a state with Poissonian statistics}

Given the outcomes of a joint measurement of a two-mode state $\rho$, the joint probability distribution is given by $p(m_1,m_2)={\rm tr}(\rho\Pi_{m_1}\otimes \Pi_{m_2})$. Moments of this distribution, $\braket{\widehat{m_1^p}\widehat{m_2^q}}$ can be calculated via the $p$-moment 
operator of the photocount,
\begin{eqnarray}\label{pmoment}\nonumber
\widehat{m^p}&=& \sum_{m=0}^{N} m^p \Pi_m\\
&=& N^p I - \sum^{N-1}_{m=0}\sum^\infty_{n=m} (N^p-m^p) w_{m,n}(\eta)|n
\rangle \langle n| \ .
\end{eqnarray}
We note that since they are the operatorial moments of a 
POVM, $\widehat{m^p}\neq\widehat{m}^p$ \cite{jpdparis} unless the outcomes $\Pi_m$ are pairwise orthogonal which is not the case here. The first two moment 
operators are
\begin{eqnarray}
\label{first}\widehat{m}&=&\eta \widehat{n}-\sum^\infty_{n=N+1} C_{n}(N,\eta)
\ket{n}\bra{n} \ , 
\end{eqnarray}
where $\hat{n}$ is the number operator, and
\begin{eqnarray}
\widehat{m^2} = &\eta^2& \widehat{n}^2 +\eta(1-\eta)\widehat{n} - (2 N +1) \sum_
{n=N+1}^\infty C_n(N, \eta)|n \rangle \langle n|+ 2 \sum_{n=N+2}^
\infty  D_n(N,\eta) |n\rangle \langle n|
\end{eqnarray}

\begin{eqnarray}
\nonumber C_n(N,\eta)&\equiv&\left (\begin{array}{c}n \\ N+1\end{array}\right ) 
x^{N+1}(1-\eta)^n F(2, N-n+1, N+2|-x) \ ,\\
\label{last}D_n(N,\eta)&\equiv&\left (\begin{array}{c}n \\ N+2\end{array}\right )
x^{N+2} (1-\eta)^n F(3, N-n+2, N+3|-x) \ ,
\end{eqnarray}
with $x=\eta/(1-\eta)$, and $F(a,b,c|x)$ is Gauss's hypergeometric function.
 
\begin{figure}
\begin{center}
\includegraphics[width=3in]{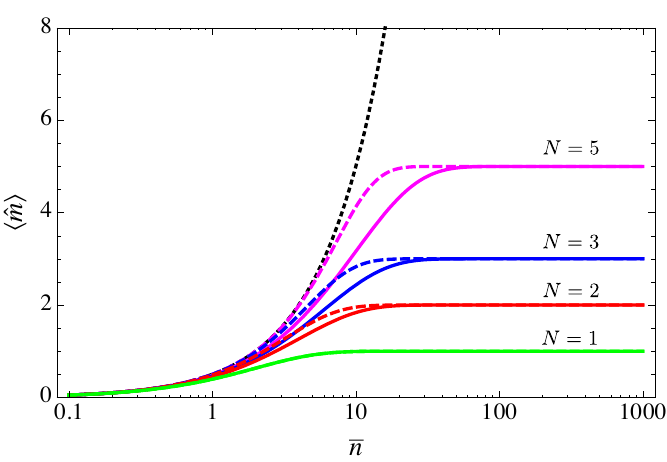}
\caption{A plot of the detected photocount $\braket
{\widehat{m}}$ with the average number of photons $
\bar{n}$ in the incident beam for the lossy detector described by the POVM 
models (a) $\{\Pi_m\}$ of Eqns.~(\ref{lossy})--(\ref{N}) (solid), and (b) 
$\{\Pi_m^{\rm (sv)}\}$ of Eqn.~(\ref{SVmodel}) (dashed)
 with various maximum photocount of the detector, $N$, and 
quantum efficiency $\eta=0.5$. Quantum efficiency 
which is the ratio of the detected 
photocount to the incident photon number, i.e. the slope of 
the detected photon curve, tends to 0 at large 
$\bar{n}$. The response of a detector with just loss is also shown (black dotted) for comparison.}
\label{sat_curves}
\end{center}
\end{figure}

Without saturation, the first- and second- moment operators are $\widehat{m}_0=\eta 
\widehat{n}$ and $\widehat{m_0^2}=\eta^2 \widehat{n}^2 +\eta (1-\eta) \widehat{n}
$, respectively. The variance of the photocounts is then $\sigma^2(m)=\braket
{\widehat{m^2}}-\braket{\widehat{m}}^2=\eta^2\sigma^2(n)+\eta(1-\eta)\braket
{\widehat{n}}$. It is clear from Eqns. (\ref{first})--(\ref{last}) that the saturation adds extra summation terms into the moment operators. For a state with Poissonian statistics we have
 \begin{eqnarray}\label{sat_curve}\nonumber
 \braket{\widehat{m}}_{\textsc{p}} &=& N - [N e_{N-1}(\eta\bar{n})-\eta\bar{n}e_{N-2}
 (\eta\bar{n})]e^{-\eta \bar{n}} \ , \\
\label{sat_curve2}  \braket{\widehat{m^2}}_{\textsc{p}} &=& N^2 - \frac{(\eta\bar{n})^N(N+\eta
\bar{n})e^{-\eta\bar{n}}}{\Gamma(N)}+[(\eta\bar{n})^2+\eta\bar{n}-N^2]e^{-\eta
\bar{n}}e_{N-1}(\eta\bar{n}) \ ,
\end{eqnarray}
where $e_n(x)$ is the exponential sum function
\begin{eqnarray}
e_n(x)=\sum^n_{k=0}\frac{x^k}{k!} \ .
\end{eqnarray}
The dependence of the average photocount of a single detector on the average photon number of an impinging Poissonian light is shown in Figure \ref{sat_curves}.

When $\bar{n}\rightarrow \infty$, the average number of photocounts tends to the saturation value $N$ with the following behavior:
\begin{eqnarray}\label{asymptote}
\braket{\widehat{m}}_{\rm p}&=&N +\mathcal{O}\left(\frac{1}{\bar{n}^2}\right) - 
e^{-\eta\bar{n}}\bar{n}^N\left(\frac{\eta^{N-1}}{(N-1)!\bar{n}}-\mathcal{O}\left
(\frac{1}{\bar{n}^2}\right)\right)\ .
\end{eqnarray}
By contrast when $\bar{n}\ll N$, the saturation effect is negligible so the 
expected detected photons behave like those without saturation,
\begin{eqnarray}\label{firsto}
\braket{\widehat{m}}_{\rm p} = \eta \bar{n} +\mathcal{O}\left(\bar{n}^2\right)-
\bar{n}^N\left(\frac{\eta^{1+N}\bar{n}}{(N+1)!}-\mathcal{O}(\bar{n}^2) \right)\ .
\end{eqnarray}
When $(\eta\bar{n})^N\ll (N+1)!$, then $\braket{\widehat{m}}_{\textsc{p}} =
\eta \bar{n}$ which identifies the linear regime of the detector. The POVM in Eqns.~(\ref{lossy})-(\ref{N}) describes the average photocount of a detector expected in the asymptotic limit of small and large average impinging photon number.

For comparison, the average photocount of the detector given by the Sperling-Vogel-Agarwal model is also plotted in Figure \ref{sat_curves}. From these plots, we can see that this model also gives a detector that saturates at the maximum resolvable photocount, $N$. Its behavior, even though motivated by a different physical situation, is qualitatively similar to that of the POVM $\{\Pi_m\}$. As such, we proceed with our analysis of the variance of difference using only the POVM $\{\Pi_m\}$.

\section{Variance of difference}
Given the two photocounts from the joint detection, their difference is
\begin{eqnarray}\label{D}
\widehat{D}=\widehat{m}_1-\widehat{m}_2 \ ,
\end{eqnarray}
with integer eigenvalues $d=0,\pm 1,\pm 2, \ldots$. The POVM of the measurement of the difference photocount, $\Theta_d$, is
\begin{eqnarray}
\Theta_d=\sum^{N-|d|}_{q=0} \left\{\begin{array}{cc} \Pi_{q
+d}\otimes\Pi_q \ &, \ d>0 \\[1 ex] \Pi_{q}\otimes
\Pi_q \ &, \ d=0\\[1 ex] \Pi_{q}\otimes\Pi_{q-d}\ &, \ 
d<0  \end{array} \right. \ ,
\end{eqnarray}
where $\Pi_m$ is the $m$-th POVM outcome of Eqn. (\ref{lossy}). The moments of 
the difference photocount distribution are then
\begin{eqnarray}\nonumber
\widehat{D} &=& \sum_d d \Theta_d = \widehat{m}_1 - \widehat{m}_2 \ , \\
\widehat{D^2} &=& \sum_d d^2 \Theta_d = \widehat{m_1^2}+\widehat{m_2^2} -2\widehat{m}_1\widehat{m}_2\ .
\end{eqnarray}
The VDP for a given quantum state is $\sigma^2(d)=
\braket{\widehat{D^2}}-\braket{\widehat{D}}^2$. Let us find the VDP for the TMC and the TWB states. For the TMC state, we get
\begin{eqnarray}\label{vodcoh}
\sigma^2_{\alpha}(d)=\sigma^2_{\alpha}(m_1)+\sigma^2_{\alpha}(m_2) \ ,
\end{eqnarray}
where $\sigma^2_{\alpha}(m_j)=\braket{\widehat{m_j^2}}_{\rm p}-\braket{\widehat
{m_j}}_{\rm p} ^2$. For the TWB state, the quantum correlations give nonzero terms in the covariance of the joint photostatistics, and this shows up in the VDP,
\begin{eqnarray}\label{vodtwb2}
\sigma^2_{X}(d)&=&\sigma^2_{\alpha}(m_1)+\sigma^2_{\alpha}(m_2)- 
2\braket{\widehat{m}_1 \widehat{m}_2}_X +2\braket{\widehat{m}_1}_{\rm p}\braket
{\widehat{m}_2}_{\rm p} \ ,
\end{eqnarray}
where

\begin{eqnarray}\label{vodtwb}
\nonumber\braket{\widehat{m}_1\widehat{m}_2}_X &=& \eta_1 \eta_2 \bar{n}(1+\bar
{n})-\sum^\infty_{n=N_2}|b_n|^2 \eta_1 n  C_n(N_2,\eta_2)-\sum^\infty_{n=N_1}|
b_n|^2 \eta_2 n  C_n(N_1,\eta_1)\\
&&+\sum_{n={\rm max}(N_1,N_2)}^\infty|b_n|^2 C_n(N_1,\eta_1)C_n(N_2,\eta_2) \ .
\end{eqnarray}

\begin{figure}
\begin{center}
\includegraphics[width=6in]{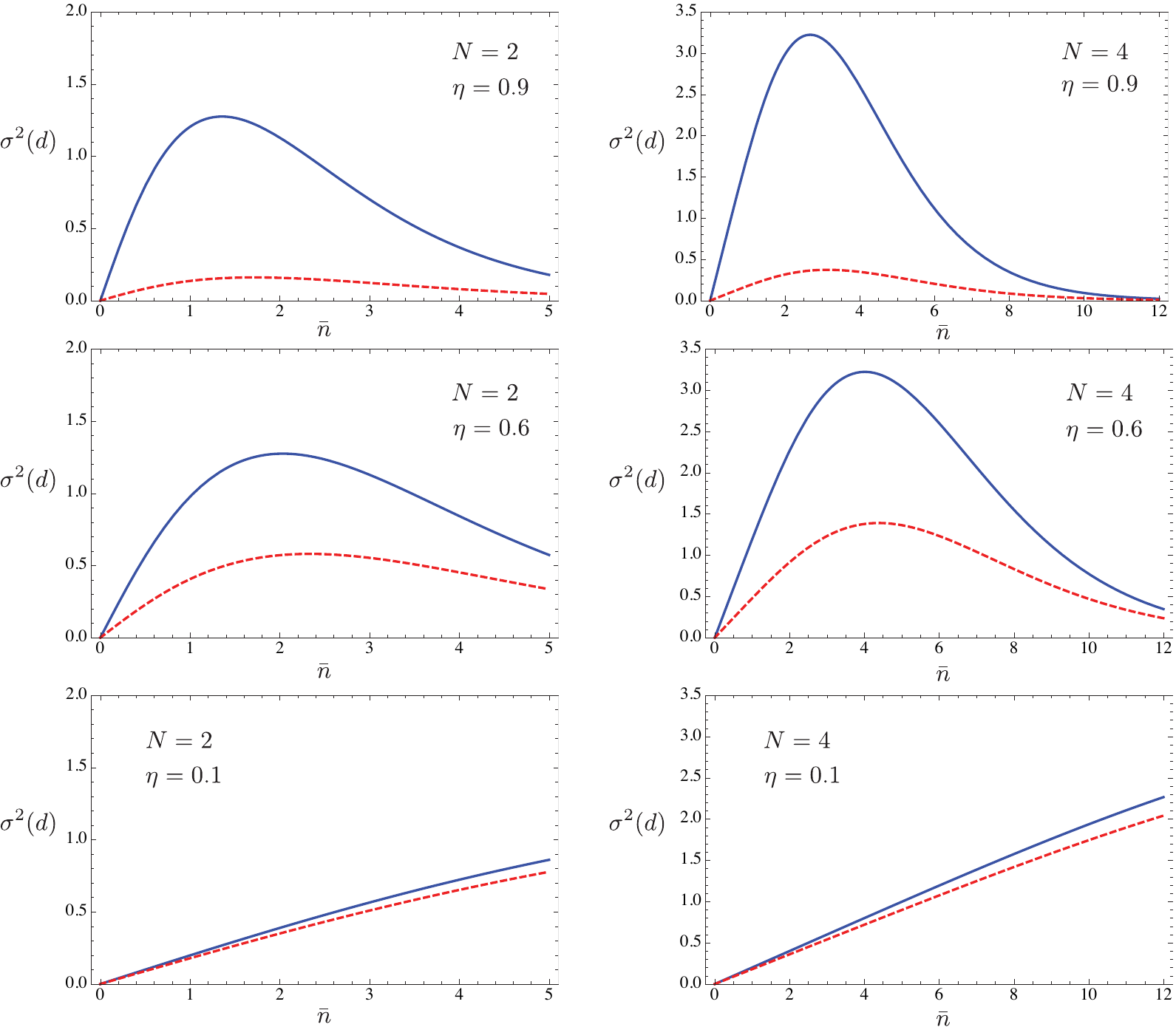}
\caption{Variance of the difference of photocounts (VDP), $
\sigma^2(d)$, as a function of the mean photon number of the input signal for 
the TMC state (blue solid line) and the TWB state (red dashed line), where $
\eta_1=\eta_2=\eta$ and $N_1=N_2=N$ with various values of $N$ and $\eta$ to 
illustrate the effects of saturation and loss.}
\label{vod}
\end{center}
\end{figure}

In Figure \ref{vod} , we show the effects of loss and saturation on measuring the VDP for the TMC and the TWB states with different values of $N_1=N_2=N$ and $\eta_1=\eta_2=\eta$. Two observations can be made. First, the saturation causes the variance to decrease with increasing average photon number for both the TMC and the TWB states and asymptotically approach zero. This means that the VDP cannot provide a reliable discrimination of classical and quantum correlations at values of $\bar{n}$ that are too large. Second, loss degrades the measurement of the quantum correlation since with increasing loss, the VDP for the TWB becomes closer to that for the TMC.

\begin{figure}
\begin{center}
\includegraphics[width=3.5in]{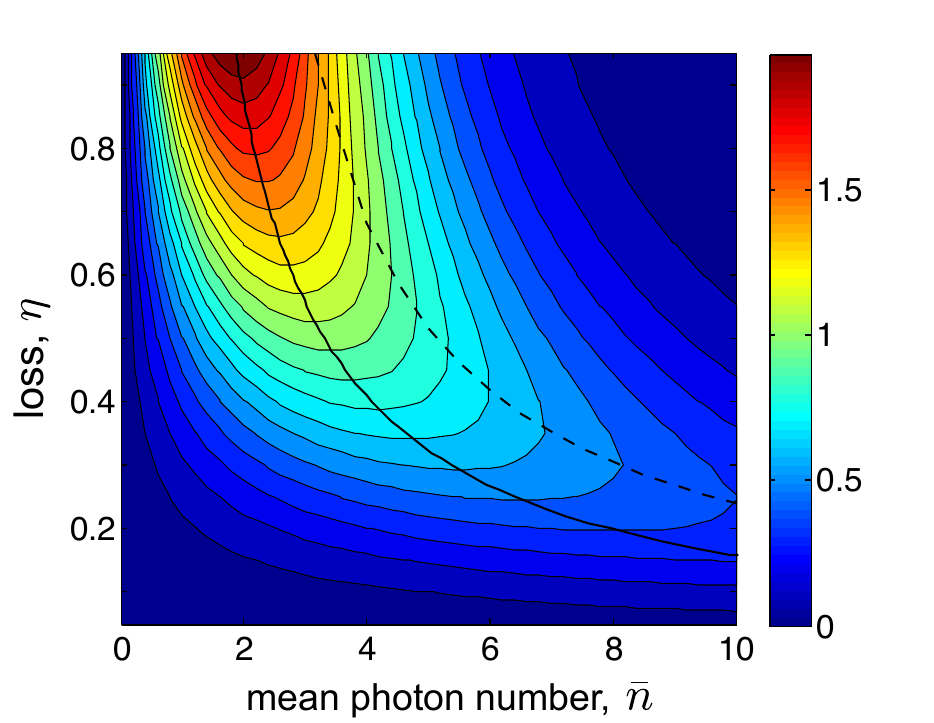}
\caption{(Color online) A contour plot of $Q$ versus mean photon number, $\bar{n}$, and loss, $\eta$, for a balanced joint detection where $\eta_1=\eta_2=\eta$ and $N_1=N_2=3$. In this figure, the maximum $Q$ value is clearly visible. The solid line shows the optimal $\bar{n}$ for a given $\eta$ and the dashed line shows the optimal $\eta$ for a given $\bar{n}$. The higher the value of $Q$, the better the discrimination between quantum and classical correlations. Note that for a fixed value of $\bar{n}$, the best discrimination does not always occur at $\eta=1$ (zero loss). A higher loss is favored at higher mean photon numbers where saturation gets more prevalent to obtain a larger $Q$ value. However, higher losses in the photocount will degrade the measurement of the VDP.}
\label{Qplot}
\end{center}
\end{figure}

However, there exists a range of $\bar{n}$ for such a discrimination to be possible. Let us define a quantity $Q$ by means of
\begin{eqnarray}\label{Q}
Q &\equiv& Q(N_1, N_2, \eta_1, \eta_2,\bar{n})=\sigma_{\alpha}^2(d)-\sigma_{X}^2(d) \ .
\end{eqnarray}
$Q$ is an indication of how good the discrimination between the 
classical and quantum correlations is. Figure \ref{Qplot} shows 
the behavior of $Q$ versus $\bar{n}$. The optimal discrimination happens for the largest $Q$ 
values  for some $\bar{n}_{\rm max}$ that we can solve for if 
we know the quantum efficiencies, $\eta$, and the maximum photocount of the detector, $N$, of the detectors. The optimal discrimination does not always occur when $\eta=1$ due to saturation. When the mean photon number is high and saturation is likely to occur, having some loss in the detectors will help to offset the effect of saturation. However, loss will degrade the quality of the measurement of the VDP. As such, there is an interplay between loss and saturation.

The analytic forms for the VDP for the TMC and TWB states measured by a PNRD with saturation and loss in Eqns.~(\ref{vodcoh})--(\ref{vodtwb}) are the main results of the paper. Using these results, we analyzed the VDP for different values of $\eta$ and $N$ thus illustrating the significance of loss and saturation to the measurement of the VDP. Thus, such an analysis is helpful for interpreting experimental data. We will now show in the next section how we can use the features of saturation in the analytic model in the absolute calibration of the PNRDs.

\section{Absolute Calibration of PRNDs using the NRF}

Sometimes it is helpful to consider a related measure of quantum correlation---the NRF,
\begin{eqnarray}
{\rm NRF}=\frac{\sigma^2(d)}{\braket{\widehat{m}_1}+\braket{\widehat{m}_2}} \ .
\end{eqnarray}
For the TMC state, ${\rm NRF}=1$ at any value of quantum efficiencies for the detectors. For the TWB, ${\rm NRF}=0$ with detectors of perfect quantum efficiencies and ${\rm NRF}=1-\frac{2\eta_1\eta_2}{\eta_1+\eta_2}$ with lossy detectors, where $\eta_1$ and $\eta_2$ are their respective quantum efficiencies. This makes the NRF a useful measure of quantum correlations for two-mode states, as well as a calibration measure for the quantum efficiencies of detectors \cite{Agafonov_cal}. We will show here how one can achieve absolute calibration, which is a calibration process without the use of a reference detector. First, we will describe the current experimental procedure for absolute calibration which is possible in the linear regime of a PNRD. Then using the analytic form for the photocount, a range of values of average photon number for the linear regime can be derived. Second, we will suggest the use of the analytic form in Eqn. (\ref{vodcoh})--(\ref{vodtwb}) for absolute calibration beyond the linear regime of the PNRD.

For the first method, let us consider the scenario in which we are given two detectors 
with unknown quantum efficiencies and saturation characteristics. 
Absolute calibration, for example, can be achieved if we have a TWB state by measuring ${\rm NRF}$ in the linear 
regime of the detectors. The calibration routine consists of these three steps:
\begin{enumerate}
\item Measure the photocounts, $\braket{\widehat{m}_1}$ and $\braket{\widehat{m}
_2}$, of the two detectors with increasing $\bar{n}$. This can be 
done by increasing the pump power of the laser. From these data, the 
mean photocounts, VDP, and 
consequently the NRF, can be calculated.

\item \label{k}Let us define the ratio of the two quantum 
efficiencies, $k=\frac{\eta_1}{\eta_2}$. In the regime where the photocounts are approximately linear with respect to $\bar{n}$, then, $\frac{\braket{\widehat{m}_1}}{\braket{\widehat{m}_2}}\approx k$. So $k$ can be found by calculating the ratios of the two photocounts.
\item For this range of average photon number, ${\rm NRF} = 1-\frac{2 \eta_1}{1+k}$. Using the measured NRF and the value of $k$ found in Step \ref
{k}, $\eta_1$ can be calculated. Finally, we can calculate $\eta_2$ from the 
values of $k$ and $\eta_1$ found. 
\end{enumerate}
Thus, absolute calibration of the detector is possible in the linear regime of the detector. This is quantified by the range of $\bar{n}$ in which the term linear in $\bar{n}$ is much larger than the next order term in Eqn. (\ref{sat_curve}),
\begin{eqnarray}\label{regime}
\frac{(\eta\bar{n})^N}{(N+1)!}\ll 1 \ .
\end{eqnarray}
The higher the saturation point is, the broader the linear regime of the detector. Although it is difficult to pinpoint exactly how small the value on the left-hand side of Eqn. (\ref{regime}) must be, current experiments suggest that values of about $0.1$ are sufficient \cite{migdall}. The worst range of values of $\bar{n}$ for the linear regime is given for $N=1$, where we would require that $\bar{n}<\frac{0.1}{\eta}$. For an APD where $N=1$, the quantum efficiency is typically around $33\%$ \cite{Kok}, thus the linear regime is where $\bar{n}< 0.6$. However, for a VLPC with a typical saturation value of $N=10$ and quantum efficiency $\eta=85\%$ \cite{Kok}, we have $\bar{n}< 5.4$. So the linear regime of the VLPC is about nine times that of the APD in this case. This puts a quantitative estimate for the linear regime of a PNRD from its characteristics which are typical for that type of detector and can be taken from its manual. Once an approximate linear regime is determined, a careful calibration can be done in this range of average photon numbers.

In practice, using the analytic forms in Eqns. (\ref{sat_curve}), (\ref{vodcoh}) and (\ref{vodtwb}), we can achieve absolute calibration beyond the linear regime of the PNRD with just a TMC state using the following protocol:
\begin{enumerate}
\item Split a pulsed laser beam with a 50-50 non-polarizing beam splitter into two parts with equal intensities to form a TMC state. One has to make sure that the laser produces a coherent state which is free of any additional classical noise, see \cite{BR}.

\item Measure the photocounts of the beam in each arm with a detector where the detection intervals are triggered by the laser. Then increase $\bar{n}$ by increasing the pump power until there is no visible change in the mean photocount with increasing $\bar{n}$. In this limit, the detectors are saturated and we have $N_1=[\braket{\widehat{m}_1}_{\rm sat}]$ and $N_2=[\braket{\widehat{m}_2}_{\rm sat}]$, where the subscript ``sat" indicates values at saturation and $[\cdot]$ denotes the closest integer.
\item Calculate the difference of photocounts for each detection interval and then the variance of this difference for each $\bar{n}$ value.
\item Using the found values of $N_1$ and $N_2$ we can generate an equation for the NRF of the TMC state as measured by the PNRD from Eqns. (\ref{sat_curves}), (\ref{vodcoh}) and (\ref{vodtwb}). A nonlinear fit of the experimental data of the VDP for the TMC state can then be done with the analytic form for the ${\rm NRF}$ using a suitable numerical algorithm. This will yield fitted values of $\eta_1$ and $\eta_2$ thus completing the calibration of the PNRDs.
\end{enumerate}
This absolute calibration procedure with the TMC state is possible owing to the features of the VDP in the presence of saturation. It would not be possible if detectors operating in the linear regime or detectors with just losses were used, because in those cases, the VDP would be insensitive to the quantum efficiencies of the detectors. The advantage of using this second method instead of the first one is that we do not have to rely on a TWB state which is harder to generate than a TMC state. The disadvantage is that we have to use a numerical nonlinear fit method for the calibration which might yield more than one reasonable fit and the errors in these fits are generally larger than those for a linear regression fit required for the first method.

\section{Conclusion}
A model of joint photodetection of lossy detectors with saturation is presented. Using this model, we have derived an analytic form for the VDP which is tractable for calculations. We then used this analytic form to study the effects of saturation on measurements of the VDP. We found that saturation diminishes this measure, and the variance of photocounts of the TMC and TWB states cannot be distinguished at large mean photon numbers. Loss in the PNRDs degrades the quality of the measurement of the quantum correlation. On the other hand, loss can offset the effects of saturation at high average photon numbers and the VDP is still a good measure of quantum correlations as long as we are not far from the optimal $Q$.

Saturation also limits the range of $\bar{n}$ in which data is useful for calibration. This has been known for some time experimentally, but here we have derived quantitatively what this range is in terms of relative values of quantum efficiency and saturation value. With the analytic forms for the measured photocounts and VDP, we can do an absolute calibration with a TMC state. The generation of a TMC state is far simpler than that of a TWB state where the latter is required for absolute calibration.

As the focus of this paper is on loss and saturation effects in PNRDs, we have not looked at possible extensions of the measurement model. For example, apart from loss and saturation, PNRDs can also suffer from dark counts and crosstalk effects \cite{crosstalk}, both of which add spurious counts to the data. In using the measurement model for accounting for noise in actual experiments, one would have to take these effects into account. This can be done by modifying Eqn. (\ref{lossy}) with the appropriate conditional probability matrices representing different noise types. Another useful extension of the measurement model is that of a variable saturation value, $N$. Here, we have considered a fixed saturation value but, in practice, a PNRD can have different saturation values at different intensities or wavelengths of incident light. A more robust measurement model can be obtained by taking this variability into account.

\section*{Acknowledgements}
We would like to thank Hui Khoon Ng, Matteo G.A. Paris, Bj\"orn Hessmo and Alexander Ling for helpful discussions. This work was supported by the A*STAR Investigatorship grant as well as the Ministry of Education, partly through the Academic Research Fund [Tier 3 MOE2012-T3-1-009], and the Singapore National Research Foundation (NRF) [grant number NRF-NRFF2013-01].


\begin{thebibliography}{29}
\providecommand{\natexlab}[1]{#1}

\bibitem[1]{noon}
Afek, I.; Ambar, O.; Silberberg, Y.~High-NOON states by mixing quantum and
  classical light.  {\em Science}  {\bf 2010}, {\em 328}, 879.

\bibitem[2]{pnc}
Bondani, M.; Allevi, A.; Zambra, G.; Paris, M.G.A.; Andreoni, A.~Sub-shot-noise
  photon-number correlation in a mesoscopic twin beam of light.  {\em Phys.
  Rev. A}  {\bf 2007}, {\em 76}, 013833.

\bibitem[3]{litho}
Boto, A.N.; Kok, P.; Abrams, D.S.; Braunstein, S.L.; Williams, C.P.; Dowling, J.P.~Quantum Interferometric Optical Lithography: Exploiting Entanglement to Beat
  the Diffraction Limit.  {\em Phys. Rev. Lett.}  {\bf 2000}, {\em 85},
  2733--2736.

\bibitem[4]{metrology}
Giovannetti, V.; Lloyd, S.; Maccone, L.~Quantum Metrology.  {\em Phys. Rev.
  Lett.}  {\bf 2006}, {\em 96}, 010401.

\bibitem[5]{cryptography}
Ekert, A.K.~Quantum Cryptography based on Bell's Theorem.  {\em Phys. Rev.
  Lett.}  {\bf 1991}, {\em 67}, 661--663.

\bibitem[6]{migdall}
Eisaman, M.D.; Fan, J.; Migdall, A.; Polyakov, S.V.~Single-photon sources and
  detection.  {\em Rev. Sci. Instrum.}  {\bf 2011}, {\em 82}, 071101.

\bibitem[7]{PNRD}
Hadfield, R.H.~Single-photon detectors for optical quantum information
  applications.  {\em Nature Photonics}  {\bf 2009}, {\em 3}, 696--705.

\bibitem[8]{cryo1}
Kim, J.; Takeuchi, S.; Yamamoto, Y.; Hogue, H.H.~Multiphoton detection using visible
  light photon counter.  {\em Appl. Phys. Lett.}  {\bf 1999}, {\em 74},
  902--904.

\bibitem[9]{cryo2}
Rosenberg, D.; Lita, A.E.; Miller, A.J.; Nam, S.W.~Noise-free high-efficiency
  photon-number-resolving detectors.  {\em Phys. Rev. A}  {\bf 2005}, {\em 71},
  061803(R).

\bibitem[10]{TMD}
Fitch, M.J.; Jacobs, B.C.; Pittman, T.B.; Franson, J.D.~Photon-Number Resolution using
  Time-Multiplexed Single-photon detectors.  {\em Phys. Rev. A}  {\bf 2003},
  {\em 68}, 043814.

\bibitem[11]{Silberberg}
Afek, I.; Natan, A.; Ambar, O.; Silberberg, Y.~Quantum State measurements using
  multipixel photon detectors.  {\em Phys. Rev. A}  {\bf 2009}, {\em 79},
  043830.

\bibitem[12]{statistics1}
Lee, H.; Yurtsever, U.; Kok, P.; Hockney, G.M.; Adami, C.; Braunstein, S.L.; Dowling, J.P.~Towards photostatistics from photon-number discriminating detectors.
  {\em J. Mod. Opt}  {\bf 2004}, {\em 51}, 1517--1528.

\bibitem[13]{dt1}
Feito, A.; Lundeen, J.S.; Coldenstrodt-Ronge, H.; Eisert, J.; Plenio, M.B.; Walmsley, I.A.~Measuring Measurement: Theory and Practice.  {\em N. J. Phys.}  {\bf
  2009}, {\em 11}, 093038.

\bibitem[14]{dt2}
Lundeen, J.S.; Feito, A.; Coldenstrodt-Ronge, H.; Pregnell, K.L.; Silberhorn,
  Ch.; Ralph, T.C.; Eisert, J.; Plenio, M.B.; Walmsley, I.A.~Tomography of Quantum
  Detectors.  {\em Nature Phys.}  {\bf 2009}, {\em 5}, 27--30.

\bibitem[15]{dt3}
D'Ariano, G.M.; Maccone, L.; Presti, P.L.~Quantum Calibration of Measurement
  Instrumentation.  {\em Phys. Rev. Lett.}  {\bf 2004}, {\em 93}, 250407.

\bibitem[16]{Ramilli}
Ramilli, M.; Allevi, A.; Chmill, V.; Bondani, M.; Caccia, M.; Andreoni, A.~Photon-number Statistics with Silicon Photomultipliers.  {\em J. Opt. Soc.
  Am. B}  {\bf 2010}, {\em 27} (5), 852--862.

\bibitem[17]{Kalashnikov}
Kalashnikov, D.A.; Tan, S.H.; Chekova, M.V.; Krivitsky L.A.~Accessing Photon Bunching
  with a Photon Number Resolving multi-pixel detector.  {\em Opt. Express}
  {\bf 2011}, {\em 19} (10), 9352--9363.

\bibitem[18]{Gisin}
Eraerds, P.; Legr$\acute{e}$, M.; Rochas, A.; Zbinden, H.; Gisin, N.~SiPM for fast
  photon-counting and multiphoton detection.  {\em Opt. Express}  {\bf 2007},
  {\em 15} (22), 14539--14549.

\bibitem[19]{Kok}
Kok, P.; Lovett, B.W.~Introduction to Optical Quantum Information Processing.
  Cambridge University Press: Cambridge, UK, 2010.

\bibitem[20]{SVA2012}
Sperling, J.; Vogel, W.; Agarwal, G.S.~True photocounting statistics of
  multiple on-off detectors.  {\em Phys. Rev. A}  {\bf 2012}, {\em 85}, 023820.

\bibitem[21]{intense}
Gl{\"o}ckl, O.; Anderson, U.L.; Leuchs, G.~Verifying continuous-variable
  entanglement of intense light pulses.  {\em Phy. Rev. A}  {\bf 2006}, {\em 73}, 012306.

\bibitem[22]{jpdparis}
Agliati, A.; Bondani, M.; Andreoni, A.; deCillis, G.; Paris, M.G.A.~Quantum and
  Classical Correlations of Intense beams of light investigated via Joint
  Photodetection.  {\em J. Opt. B: Quantum Semiclass. Opt.}  {\bf 2005}, {\em 7},
  S652--S663.

\bibitem[23]{Klyshko}
Klyshko, D.N.~Use of two-photon light for absolute calibration of photoelectric
  detectors.  {\em Sov. J. Quantum Electron}  {\bf 1980}, {\em 7} (9), 1112--1116.

\bibitem[24]{cali}
Brida, G.; Genovese, M.; Gramegna, M.~Twin-photon techniques for photo-detector
  calibration.  {\em Laser Phys. Lett.}  {\bf 2006}, {\em 3}, 115--123.

\bibitem[25]{Agafonov_cal}
Agafonov, I.N.; Chekhova, M.V.; Iskhakov, T.Sh.; Penin, A.N.; Rytikov, G.O.; Shumilkina, O.A.~Absolute
  calibration of photodetectors: photocurrent multiplication versus
  photocurrent subtraction.  {\em Opt. Lett.}  {\bf 2011}, {\em 36}, 1329-1331.

\bibitem[26]{waks}
Waks, E.; Sanders, B.C.; Diamanti, E.; Yamamoto, Y.~Highly nonclassical photon
  statistics in parametric down-conversion.  {\em Phys. Rev. A}  {\bf 2006},
  {\em 73}, 033814.

\bibitem[27]{haderka}
Haderka, O.; Pe\v{r}ina, J., Jr.; Hamar, M.; Pe\v{r}ina, J.~Direct measurement and
  reconstruction of nonclassical features of twin beams generated in
  spontaneous parametric down-conversion.  {\em Phys. Rev. A}  {\bf 2005}, {\em
  71}, 033815.

\bibitem[28]{GaussianCV}
Ferraro, A.; Olivares, S.; Paris, M.G.A.~Gaussian states in continuous variable
  quantum information.  {\em Preprint}  {\bf 2005}, {\em quant-ph/0503237v1}.

\bibitem[29]{supercondcal}
Brida, G.; Ciavarella, L.; Degiovanni, I.P.; Genovese, M.; Lolli, L.; Mingolla, M.G.; Piacentini, F., Rajteri, M.; Taralli, E.; Paris, M.G.A.~Quantum characterization of superconducting photon counters.  {\em New J.~of Phys.}  {\bf 2012}, {\em 14}, 085001.

\bibitem[30]{KT2012}
Kalashnikov, D.A.; Tan, S.-H.; Iskhakov, T.S.; Chekhova, M.V.; Krivitsky, L.A.~Measurement of two-mode squeezing with photon number resolving multipixel detectors.  {\em Opt. Lett.}  {\bf 2012}, {\em 37}, 2829-2831.






\bibitem[31]{BR}
Bachor, H.; Ralph, T.C.~A Guide to Experiments in Quantum Optics.2nd;
  Wiley-VCH: Weinheim, Germany, 2004.

\bibitem[32]{crosstalk}
Kalashnikov, D.A.; Tan, S.-H.; Kritvitsky, L.A.~Crosstalk calibration of multi-pixel photon counters using coherent states.  {\em Opt. Express}  {\bf 2012},
  {\em 20} (5), 5044--5051.


\end{thebibliography}
\end{document}